\documentstyle[prd,aps,preprint,floats,epsfig]{revtex}
\newcommand{\beq}{\begin{equation}}
\newcommand{\eeq}{\end{equation}}
\newcommand{\beqr}{\begin{displaymath}}
\newcommand{\eeqr}{\end{displaymath}}
\newcommand{\beqa}{\begin{eqnarray}}
\newcommand{\eeqa}{\end{eqnarray}}
\newcommand{\beqar}{\begin{eqnarray*}}
\newcommand{\eeqar}{\end{eqnarray*}}
\newcommand{\al}{\alpha}

\newcommand{\del}{\delta}
\newcommand{\D}{\Delta}

\renewcommand{\L}{\Lambda}

\newcommand{\w}{\omega}

\newcommand{\cR}{{\cal R}}

\newcommand{\ssc}{\scriptscriptstyle}

\newcommand{\ie}{{\it i.e.}}

\newcommand{\non}{\nonumber}

\newcommand{\ti}[1]{\widetilde{#1}}

\newcommand{\mt}[1]{\textrm{\tiny #1}}
\newcommand{\bk}[1]{#1}

\newcommand{\GC}{G_{\ssc{D}}}

\begin{document}

\thispagestyle{empty}
\rightline{\small hep--th/0107034 \hfill McGill/00-14}\nopagebreak
\vskip -.6ex
\vspace*{2cm}

\begin{center}
{\bf Brane world sum rules and the ${\rm AdS}$ soliton$^{}$\footnote{Based on a talk presented by F.~Leblond at the Eight
International Symposium on Particles, Strings and Cosmology (PASCOS 2001), University of North Carolina at Chapel Hill.}}
\vspace*{1cm}

Fr\'ed\'eric Leblond,$^{}$\footnote{E-mail: fleblond@hep.physics.mcgill.ca}
Robert C. Myers$^{}$\footnote{E-mail: rcm@hep.physics.mcgill.ca} and 
David J. Winters$^{}$\footnote{E-mail: winters@hep.physics.mcgill.ca} 
\vspace*{0.3cm}

{\it Department of Physics, McGill University\\ 3600 University Street,
Montr\'eal, QC, H3A 2T8, Canada}\\
\vspace{2cm}
ABSTRACT
\end{center}
We consider ``brane world sum rules'' for compactifications involving
an arbitrary number of spacetime dimensions. One of the most striking results
derived from such consistency conditions is the necessity for negative tension
branes to appear in five--dimensional scenarios. We show how this result
is evaded for brane world models with more than five dimensions. As an example,
we consider a novel realization of the Randall--Sundrum scenario in six
dimensions involving only positive tension branes. A complete account of our 
results appeared in hep--th/0106140.

\vfill
\setcounter{page}{0}
\setcounter{footnote}{0}
\newpage

\section{Introduction}
It is not a priori excluded that effective theories directly related to the brane world idea might reside on some corner
of the moduli space of string/M--theory. Attempts to derive phenomenologically interesting 
warped compactifications of the latter
were presented at this conference\cite{kachru}. Our approach to the problem is of the bottom--up kind though.  
In recent years, considerable efforts have been invested in the study of localized gravity using 
five-dimensional general relativity 
(see, for example, Refs. \cite{RS1,RS2}). However,
Gibbons et al.\cite{gibbons} were
able to show that negative tension branes are a necessary ingredient for a very wide class
of five-dimensional warped compactifications.
This is a rather disappointing conclusion as negative tension branes are 
inherently unstable objects. Here we generalize the work of Ref.~\cite{gibbons} to
find consistency conditions or ``sum rules'' that must be satisfied by brane world
models with an arbitrary number of spacetime dimensions.
This allows us to show that the necessity of introducing negative tension branes
is only an artifact of five-dimensional spacetime physics.
As an application, we present a six--dimensional realization of the Randall--Sundrum model
based on the geometry of the ${\rm AdS}$ soliton. A complete account of this
work appears in Ref. \cite{myers}.
    
\section{Consistency conditions for brane worlds in arbitrary dimensions}
The ansatz we use for the brane world consists 
of the $D$-dimensional warped product metric of a $(p+1)$-dimensional spacetime
with coordinates $x^{\mu}$ and a $(D-p-1)$-dimensional compact internal space with coordinates $y^{m}$,
\beq
\label{metric}
ds^2 = G_{{\ssc MN}}(X)dX^{{\ssc M}}dX^{{\ssc N}} = g_{mn}(y)dy^mdy^n+W^2(y)g_{\mu\nu}(x)dx^\mu dx^\nu.
\eeq 
The stress-energy tensor generating this configuration is assumed to have the following
simple form,
\beq
\bk{T}_{{\ssc MN}}=-\frac{\Lambda G_{{\ssc MN}}}{8\pi \GC}-
\sum_{\ssc{i}} T_{\ssc{q}}^{\ssc{(i)}} P[G_{\ssc{MN}}]_{\ssc{q}}^{\ssc{(i)}}
\D^{\ssc{(D-q-1)}}(y-y_{\ssc{i}}),\label{stressans}
\eeq
which contains a cosmological constant and a collection of branes of various dimensions.
To simplify the presentation here, we have not included any matter fields in the bulk or
on the branes --- see, however, Ref.~\cite{myers}. The $i^{\mt{th}}$
brane is a $q$--brane ($p\le q\le D-2$) with tension $T_{\ssc{q}}^{\ssc{(i)}}$ (units of
energy/length$^q$) and transverse
coordinates $y_{\ssc{i}}$. $P[G_{\ssc{MN}}]_{\ssc{q}}^{\ssc{(i)}}$ is the pull--back of the bulk metric to
the worldvolume of the $q$--brane. In this ansatz, 
$\D^{\ssc{(D-q-1)}}(y-y_{\ssc{i}})$ denotes
that covariant combination of delta functions and (geo)metric factors necessary
to position the brane.  Note that implicitly we are assuming that all of the 
branes are extended in the $x^\mu$ directions, and if $q>p$ for
a particular brane, it spans a $(q-p)$--cycle in the internal space.

Substituting Eqs.~(\ref{metric}) and (\ref{stressans}) into Einstein's equations,
we can derive a one--parameter ($\alpha$) family of consistency equations by integrating
over the compact internal space\cite{myers},
\beqa
\label{intcond}
&&\oint W^{\al+1} \bigg(\al\, \cR\,W^{-2}+(p-\al)\,\ti{\cR}
-\left[\gamma+(D-p-1)\ti{\gamma}\right]\Lambda
\\
&&\qquad\qquad-8\pi \GC\displaystyle \sum_{\ssc{i}} \left(\gamma+(q-p)
\ti{\gamma}\right)T_{\ssc{q}}^{\ssc{(i)}}\D^{\ssc{(D-q-1)}}(y-y_{\ssc{i}})
\bigg)\ =\ 0,
\non
\eeqa
where $\cR$ and $\ti{\cR}$ denote the Ricci scalars of the ``brane'' and internal space metrics,
$g_{\mu\nu}$ and $g_{mn}$, respectively. We have also introduced the constants:
\begin{eqnarray}
\gamma=\frac{p+1}{D-2}\left[(p-2\al)(D-p-1)+2\al\right],\qquad\ti{\gamma}=\frac{p(2\al-p+1)}{D-2}.  \nonumber
\end{eqnarray}
Note that Eq.~(\ref{intcond}) is merely a convenient
re-expression of some components of the Einstein equations as an infinite set of consistency equations relating the geometry of the brane world
to its stress-energy content. To gain some insight into these
consistency equations, we consider the phenomenologically interesting
case $p=3$. We also make the choice $\al=-1$ since it simplifies the equation considerably by removing the warp factor from
most terms. With these choices, the constraint may be written as:
\beq
\oint \Bigg(-\cR W^{-2}+4\ti{\cR}-\frac{8(D-5)}{D-2}\Lambda \Bigg)
=\frac{32\pi \GC}{D-2}\displaystyle \sum_{\ssc{i}} (5D-13-3q)L_{\ssc{i}}\,T_{\ssc{q}}^{\ssc{(i)}}\ ,
\label{simple}
\eeq
where $L_{\ssc{i}}$ is the area of the $(q-p)$--cycle in the internal space
spanned by the $i^{\mt{th}}$ brane. If $q=p$ (\ie, the brane is not extended
in the internal space), then $L_{\ssc{i}}=1$.

If $D=5$, the consistency equation (\ref{simple}) is 
independent of $\L$, and $\ti{\cR}=0$ since the internal space is one-dimensional.
Eq.~(\ref{simple}) then reduces to the 
sum rule:
\beq
-\cR \oint  W^{-2}=32\pi \GC \sum_{i} T_3^{\ssc{(i)}}.
\eeq
Our
result here essentially reproduces that given in Ref.\cite{gibbons}. In particular, if the curvature
on the branes is positive or vanishes, we have $\sum_{\ssc{i}} T_3^{\ssc{(i)}}\le 0$
and so we must include a negative tension brane for a consistent model.

For $D=6$, Eq.~(\ref{simple}) becomes
\beq
\oint \left(-\cR W^{-2}+4\ti{\cR}-2\Lambda\right)
=8\pi \GC \displaystyle \sum_{\ssc{i}} (17-3q)L_i\,T_q^{\ssc{(i)}}\ .
\label{d6cons}
\eeq
Note that on the RHS, there are contributions coming from three--
and four--branes, all with positive coefficients. On the LHS, however,
we have additional contributions coming from the cosmological constant and
the curvature of the two--dimensional internal space. Certainly, these
contributions afford us much more leeway in constructing consistent
brane world models (even when no matter fields are present).
For instance, a positive $\ti{\cR}$ and/or negative $\L$ can produce
an overall positive contribution on the LHS which could then accommodate
the appearance of only positive tensions on the RHS. Similar contributions
from the cosmological constant and internal curvature appear in
Eq.~(\ref{simple}) for all higher dimensions $D\ge6$.
Hence the sum rules are obviously much less restrictive when we go
beyond $D=5$. 

\section{$D=6$ brane world based on the ${\rm AdS}$ soliton metric}

We now consider a novel realization of the Randall-Sundrum scenario in six dimensions based on the geometry of the
${\rm AdS}$ soliton\cite{gary} --- for related work, see Refs.~\cite{nelson,chen}.
The line element of interest
is
\beq
\label{line}
ds^2=\frac{r^2}{L^2}(\eta_{\mu\nu}dx^\mu dx^\nu+f(r)d\tau^2)
+\frac{L^2}{r^2}\frac{dr^2}{f(r)},\label{sol}
\eeq
where $f(r)=1-\w^{5}/r^{5}$ and $\mu,\nu = 0,1,2,3$.  $L$ is related to the cosmological
constant by $\Lambda=-10/L^{2}$. The warp factor is $W(r)=r/L$ and the brane metric is
flat, \ie, $g_{\mu\nu}=\eta_{\mu\nu}$. The internal geometry 
closes off at $r=\w$, and the coordinate $\tau$ is periodic with period
\beq
\Delta\tau=\frac{4\pi L^2}{5\w}\left(1-\frac{\delta}{2\pi}\right) \label{singper}.
\eeq
The spacetime exhibits a conical singularity at $r=\w$ (with a deficit angle $\delta$ in $\tau$) which
can be thought of as being generated by a three-brane.
To construct a brane world geometry with a compact internal space, we truncate the AdS
soliton at some value $r=R$ and paste two identical copies back-to-back along the 
revealed hypersurface. The $r$ and $\tau$ directions then form an internal space with spherical 
topology. The resulting metric is continuous but not differentiable at the interface. The source
of the discontinuity in the extrinsic curvature is interpreted as a bound state of positive tension branes 
at $r=R$. If we assume that the tensions add linearly, the system is composed of a four-brane with tension
\beq
T_{4}=\frac{1}{\pi \GC L}\sqrt{1-\left(\frac{\w}{R}\right)^5} > 0,
\eeq 
and a three brane smeared over the $\tau$-direction with tension
\beq
T_{3}^{{\ssc (3)}}=\frac{1}{4\pi \GC}\left[\left(\frac{\w}{R}\right)^4(2\pi-\delta)\right] > 0.
\eeq
The spacetime also contains two three-branes at both ``ends'' of the internal space ($r=\omega$).
Their tension is given by (where we assume $\del>0$):
\beq
T_{3}^{{\ssc (1,2)}}=\frac{\del}{8\pi\GC} > 0.
\eeq  

In the present case, the sum rule (\ref{simple}) reduces to
\beq
\label{alpha1}
16\pi+ V_{{\ssc 2}} |\L|=32\pi\GC\displaystyle \sum_{i=1}^3 T_3^{{\ssc (i)}}
+20\pi\GC L_{\tau} T_4 ,
\eeq
where $V_{{\ssc 2}}$ is the volume of the internal space, and
$L_{\tau}=\Delta\tau\,f(R)^{\frac{1}{2}}R/L$.
We have used the property that the integral of the internal curvature over the 
two-dimensional compact space yields a topological invariant, the Euler character 
\beq
\chi=\frac{1}{4\pi} \oint \ti{\cR} = 2,
\eeq
where the two is fixed by the spherical topology.
Eq.~(\ref{alpha1}) is satisfied for the brane tensions and the cosmological constant associated with the spacetime
introduced in this section. This consistency condition makes explicit the fact that the curvature of the internal
space (yielding the $16\pi$ in Eq.~(\ref{alpha1})) and the negative cosmological constant
contribute counterterms to positive tension branes. 
We also showed 
that conditions corresponding to other values of the $\alpha$ parameter are satisfied\cite{myers}.

The ${\rm AdS}$ soliton provides an interesting realization of the Randall-Sundrum mechanism to
generate a large hierarchy\cite{RS1}. To see this we apply the coordinate
transformation, $r(y)=\w\,\textrm{cosh}^{2/5}(5y/2L)$, which modifies
the line element (\ref{line}) such that
\beq
\frac{L^2}{r^2}\frac{dr^2}{f(r)} = dy^{2}.
\eeq    
Then the warp factor becomes
\beq
W=\frac{r}{L}=\frac{\w}{L} \cosh^{2/5} \left(\frac{5y}{2L}\right)\sim \frac{\w}{L}\exp\left(
{y\over L}\right),
\eeq
where the final approximation on the RHS applies for $y\gg L$. Hence there is a large
gravitational redshift between the 
branes at $y_{\ssc{IR}}=0$ ($r=\omega$) and those at the
interface ($y_{\ssc{UV}} = 2L/5 \;\textrm{arccosh} (R/\w)^{5/2}$). 
Hence this model easily generates a large hierarchy between physics scales at the
visible brane, taken to be one
of the two three-branes generating a conical singularity at $r=\omega$,
and the six-dimensional Planck scale. The corresponding brane world model then provides
a six-dimensional realization of the RS I scenario\cite{RS1} while including
only positive tension branes.

\section*{Acknowledgments}
FL would like to thank the organizers of PASCOS 2001 for their hospitality. This research was supported in part by NSERC of Canada 
and FCAR du Qu\'ebec.

\end{document}